\documentstyle[aps,preprint,tighten]{revtex}
\begin{document}

\title{Nucleon Resonance Effects in $pp\rightarrow pp\pi^0$
near Threshold}

\author{M.T. Pe\~na$^{1,2}$, D.O. Riska$^3$, and A. Stadler$^{1,4}$}

\address{$^1$Centro de F\'\i sica Nuclear, 1699 Lisboa, Portugal}
\address{$^2$Instituto Superior T\'ecnico, 1096 Lisboa, Portugal}
\address{$^3$Department of Physics, University of Helsinki,
00014 Finland}
\address{$^4$Departamento de F\'\i sica, 
Universidade de \'Evora, 7000 \'Evora,
Portugal}
\maketitle

%\setcounter{page} {0}
%\vspace{1cm}
%\centerline{\bf Abstract}
%\vspace{0.5cm}
\abstract{
The role of the low lying nucleon resonances beyond the $\Delta(1232)$
in the reaction $pp\rightarrow pp\pi^0$ near threshold is shown to
be numerically significant by a calculation, which takes into account the
pion re-scattering contribution described by chiral perturbation
theory and the short-range mechanisms that are implied by the
nucleon-nucleon interaction model. The intermediate
$N(1440)$ (P$_{11}$) resonance is
excited by the short-range exchange 
mechanisms, while the $N(1535)$ ($S_{11}$)
and $N(1520)$ ($D_{13}$) resonances are excited by $\eta$ and $\rho$ meson
exchange, respectively. The $P_{11}$ 
increases the calculated cross section, whereas the $S_{11}$ and $D_{13}$
resonances decrease it. The calculation takes full account of
the initial and final state interactions.}

\newpage
%\centerline{\bf 1. Introduction}
%\vspace{0.5cm}
%==================================================================
\section{Introduction}

The cross section for the reaction $pp\rightarrow pp\pi^0$ near threshold 
is exceptionally sensitive to short-range exchange mechanisms in the
two-nucleon system, because the main pion exchange term is ruled out
by isospin conservation in the two-nucleon
system  \cite{LeeR,Myhr,VanK}. Pion production on a single nucleon
under-predicts the empirical cross section  \cite{Meyer1,Meyer2,Celsius} 
by a large factor  \cite{Miller}. Short-range exchange mechanisms
that are implied by the nucleon-nucleon interaction enhance the
cross section  \cite{LeeR}. The role of the residual pion exchange 
mechanisms,
which are not necessarily small when one pion is virtual, remains
contentious as conventional phenomenological meson field theory
models and chiral perturbation theory (ChPT) disagree on the sign of the
pion exchange amplitude  \cite{Myhr,VanK,Oset,Hanhart}. 

Systematic
amplitude analysis of the reactions $pn\rightarrow pp\pi^-$
and $pp\rightarrow pn\pi^+$ indicates that the short-range
mechanisms, which enhance the (small) cross section obtained
from the single nucleon pion production mechanisms, have to
dominate over the pion exchange amplitude given by chiral
perturbation theory, because empirically
the phase of the amplitudes for
production of $S$ and $P$ wave pions has to be the 
same  \cite{Tamura}. In this situation it appears natural to
investigate the role of such other short-range mechanisms, which
should be expected to contribute to this reaction, and which
may be calculated with some degree of confidence. 

The most obvious additional
short-range mechanisms are those which involve
transition couplings between different exchanged mesons, and those
that involve excitation of intermediate virtual nucleon resonances 
by short-range exchange mechanisms. The most obvious of the former
class of effects were estimated -- and found to be non-negligible --
in Ref. \cite{VanR}. The role of intermediate $\Delta(1232)$
resonances has been investigated in Refs. \cite{VanK,Nisk}, and found
to be significant despite the threshold suppression factor. 
The role of the intermediate $N(1440)$ ($P_{11}$) resonance
excited by scalar and vector exchange mechanisms was considered
in Ref. \cite{VanR} and found to be small, although the small magnitude
of the result depended on a rather uncertain estimate for the
coupling of the effective scalar field to the $N(1440)$  \cite{Mad}.

We here consider in addition the role of the intermediate $N(1535)$
($S_{11}$) and $N(1520)$ ($D_{13}$) resonances, which form the lowest
``$P-$shell'' multiplet in the baryon spectrum. The former is excited
both by pion and -- in particular -- by $\eta-$meson exchange. 
The latter is excited by pion as well as $\rho-$meson exchange.
The
pion exchange contributions are presumably included in the ``off-shell''
ChPT $\pi N$ amplitude, and therefore do not require explicit
evaluation. The $\eta-$ and $\rho-$meson exchange contributions have
have to be derived separately. Systematic calculation of these
contributions calls for employment of a boson exchange model for
the nucleon-nucleon interaction so as to avoid additional model
dependence through coupling constants, meson propagators, and
form factors. For this reason we carry out the calculation by
using the ``Bonn-B'' potential model  \cite{Machl}.

In order to obtain an illustrative as well as quantitatively realistic 
description of the resonance contributions to the near threshold
cross section for the reaction $pp\rightarrow pp\pi^0$, we perform
the cross section calculation with full account of the 
nucleon-nucleon 
interaction in the initial and final states.
 
We performed the calculations of all contributions with
the exact kinematics, i.e., we did not use the popular ``frozen 
kinematics approximation'' in which threshold kinematics is used also at
energies above threshold. 
Calculations using this approximation, as well as with
another approximation concerning the energy dependence of the two-nucleon
T-matrix, are, however, presented for comparison. 

In addition, we take explicitly into account the kinematically determined
energy of the exchanged pion in the ChPT amplitude for the
pion exchange contribution to the pion production amplitude,
a correction term that was found to be significant in Ref.~\cite{Myhr2}.
With conventional estimates for the parameters that describe
these exchange mechanisms, we obtain a fairly satisfactory
description of the cross section for 
$pp\rightarrow pp\pi^0$
in the near-threshold region up to about 300 MeV laboratory
kinetic energy without nucleon resonance and explicit
short-range effects associated with meson transition couplings.
This differs from other similar calculations based on
other nucleon-nucleon interaction models, and reflects the
extreme sensitivity to the details of the short-range
parts of the interaction models.

The contributions from the $P_{11}$ and $S_{11}$ resonances
are small, and without further short-range
contributions would lead to somewhat too large cross section
values above 300 MeV. The smallness of the contribution
from the $P_{11}$ resonance agrees with the finding in Ref.
 \cite{VanR}. The smallness of the contribution from the
$S_{11}$ is mainly due to the smallness of the $\eta-$nucleon
coupling constant. 

Choosing positive signs for the coupling parameters for all
resonances except the
$D_{13}$, that particular resonance contributes with the
opposite sign in comparison with the other two resonances, and taking
it into account actually
improves the calculated result. To this net result one may
add the contributions from the intermediate $\Delta$ resonance
and from the short-range mechanisms associated with mesonic
transition couplings, which however, when combined amount to
a very small correction, because of their tendency for cancellation.

This article is divided into 5 sections. In section 2, a brief
review of the least contentious pion production mechanisms
that have been considered in the literature is given, including
a simplified derivation of the short-range mechanisms 
associated with the nucleon-nucleon interaction. The
resonance excitation amplitudes are derived in section 3. 
The numerical results are reported in section 4, while section
5 contains a concluding discussion.

%\vspace{1cm}
%\centerline{\bf 2. Pion production and the nuclear axial current }
%\vspace{0.5cm}
\section{Pion production and the nuclear axial current}

%\centerline{\it 2.a Chiral Symmetry Constraint}
%\vspace{0.5cm}
\subsection{Chiral Symmetry Constraint}
The absence of parity doubling of the experimental hadron
spectra at low energies implies that the approximate
chiral symmetry of $QCD$ is realized in the hidden mode, and that the
low-mass pseudoscalar meson octet $\pi,K,\eta$ has to represent the
associated Goldstone bosons. It follows that the coupling of
this meson octet to hadrons has to vanish with the four-momenta
of these mesons. The coupling of pions to a nuclear system therefore
has to have the general form
\begin{equation}{\cal L}=-
{1\over f_\pi}\partial_\mu\vec \pi\cdot\vec A_\mu,\label{e:lag}
\end{equation} 
where $\vec A_\mu$ is the axial (flavor octet) current density of the
nuclear system and $f_\pi$ is the pion decay constant 
 \cite{Weinberg,Nyman}.

The interaction (1) implies that near-threshold pion production off
nuclei is governed by the axial charge density operator, which is known
to be dominated by two-nucleon mechanisms  \cite{Kubodera,Mariana,Towner}.
In this regard, the reaction $pp\rightarrow pp\pi^0$ near threshold
forms a special case, because as the main (isospin antisymmetric) pion
exchange contribution is eliminated, the reaction is governed
by short-range mechanisms.

These short-range mechanisms fall into three categories. The first is
associated with the isospin symmetric pion exchange amplitude, which,
while of vanishingly small significance for elastic pion scattering,
may be large for the off-shell amplitude involved in pion production
off nuclear systems. The second category are the short-range 
contributions to the axial charge operator that are implied by the
nucleon-nucleon interaction model, which has to be employed to
calculate the nuclear wave functions. They are represented in
Fig.~\ref{Fig.1}. The final category are the
additional model-dependent short-range contributions that are
associated with excitation of virtual intermediate nucleon resonances
by short-range 
mechanisms (Fig.~\ref{Fig.2}), and short-range mechanisms that are
associated with mesonic transition couplings. 

%\vspace{0.5cm}
%\centerline{\it 2.b Pion exchange contribution}
%\vspace{0.5cm}
%==================================================================
\subsection{Pion exchange contribution}

The isospin symmetric pion exchange contribution to the pion
production amplitude was derived by means of ChPT in
Refs.~\cite{Myhr,VanK}, and found to yield an amplitude that
disagrees in sign with that obtained by phenomenological
meson field theory models  \cite{Oset,Hanhart}. We shall here
employ the ChPT amplitude given in Ref.~\cite{VanK}, but treat
the pion energy determined kinematically exactly as in Ref.~\cite{Myhr2}.
The higher-order loop contributions that involve two nucleons
considered in Ref.~\cite{Moalem}, will be assumed to form part
of the short-range contributions that are implied by the 
nucleon-nucleon interaction model described below.

The one-nucleon and the isospin symmetric
pion exchange contribution to the axial
charge density operators then have the expressions  \cite{LeeR}
\begin{equation}
A_0^a(1)=-f_\pi
{f_{\pi NN}\over m_\pi}\vec\sigma\cdot\vec v
\tau^a,\label{e:axial}
\end{equation}
\begin{equation}
A_0^a(\pi)=f_\pi{f_{\pi NN}\over m_\pi^2}{8\pi\lambda_1\over 
\omega_\pi}
{\vec\sigma^2\cdot \vec k_2\over m_\pi^2+\vec k_2^2-\omega_k^2}
f(\vec k^2)\tau_2^a+(1\leftrightarrow 2).\label{e:piex}
\end{equation}
Here, $(\vec k_2,\omega_k)$ is the 4-momentum of the
exchanged pion that is absorbed by nucleon 2
($\vec k_2 = \vec p_2\,' -\vec p_2$), 
and $f$ is a phenomenological form factor
that dampens out high values of the exchanged momentum. We use the
parametrization of the Bonn potential \cite{Machl},
\begin{equation}
f(\vec k^2) = \left(\frac{\Lambda_\pi^2 - m_\pi^2}
{\Lambda_\pi^2 + \vec k^2} \right)^2 \, ,\label{e:ff}
\end{equation}
with the pion cut-off mass $\Lambda_\pi = 1.3$ GeV.
Note that by the Goldberger-Treiman relation
$g_A/2f_\pi=f_{\pi NN}/m_\pi$, where $f_{\pi NN}$ is the
$\pi NN$ pseudovector coupling constant ($f_{\pi NN}\simeq 1$).

The coefficient $\lambda_1$ is determined by the $S-$wave
$\pi N$ scattering lengths $a_1, a_3$ for physical $\pi N$ scattering
near threshold as
\begin{equation}
\lambda_1=-{1\over 6}m_\pi 
\left(1+{m_\pi\over m_N}\right)(a_1+2a_3).\label{e:lam1}
\end{equation}
Here, $m_N$ is the nucleon mass.
Extant empirical values for the scattering lengths 
are consistent with $\lambda_1=0$. For the off-shell $\pi N$
scattering amplitude, appropriate to pion production associated
with pion exchange, ChPT gives the following expression for 
$\lambda_1$  \cite{Myhr,VanK}:
\begin{equation}
\lambda_1={m_\pi^3\over 4 \pi f_\pi^2}\bigg[2c_1+\left(c_2+c_3-
{g_A^2\over 8 m_N}\right){\omega_q\omega_k\over m_\pi^2}
-c_3{\vec q\cdot \vec k_2\over m_\pi^2}\bigg]\, , \label{e:lam1ch}
\end{equation}
where $\vec q$ is the momentum of the produced pion
and $\omega_k=E_2\,'-E_2$. 
The coefficients $c_j$ are determined by the empirical or
empirically extracted values for the two 
$S-$wave scattering lengths,
the $\sigma$ term, and the axial polarizability 
for $\pi N$ scattering. We here use the values $c_1=-0.87 $ GeV$^{-1}$,
$c_2=3.34 $ GeV$^{-1}$, and $c_3=-5.25 $ GeV$^{-1}$ given in 
Ref.~\cite{Meissner}.

The energy of the exchanged pion $\omega_k$ is determined kinematically
as
\begin{equation}
\omega_k={\vec k_2\cdot (\vec p_2\,'+\vec p_2)\over 2 m_N},
\label{e:pien}
\end{equation}
where $\vec p_2$ 
and $\vec p_2\,'$ are the initial and final momenta of nucleon 2
(the one which does not emit the real pion), 
although it is a common approximation to set it to half of the
energy of the produced pion. The limits of the validity of that
approximation were investigated in Ref.~\cite{Myhr2}.

%\vspace{0.5cm}
%\centerline{2.c. Short-range exchange contributions}
%\vspace{0.5cm}
%==================================================================
\subsection{Short-range exchange contributions}

The contributions to the axial exchange charge operator 
associated with the short-range components of the 
nucleon-nucleon interaction have been derived in Refs.
 \cite{LeeR,Mariana}. These contributions correspond to
the nonrelativistic limit of the nonsingular part of the
axial current 5-point function with external leg
couplings, and are colloquially
referred to as nucleon-antinucleon ``pair currents'' or
``Z-graphs'', and are represented as such
by the diagrams in Fig.~\ref{Fig.1}. The form of these is implied by the
Poincar\'e invariance of the 5-point functions  \cite{Coe}.

The numerically most significant short-range mechanisms
are those that are associated with the scalar and vector
exchange terms in the nucleon-nucleon interaction.
The former may be derived directly, without
reference to the 5-point function, in the following way.

Consider the isospin independent scalar exchange component
of the nucleon-nucleon interaction, which contains the
Fermi invariant ``$S$''. To second order in $v/c$, this
interaction takes the form
\begin{equation}
v_S^+(r) S = v_S^+(r) \left(1 - {\vec p\,^2\over m_N^2}\right)
-{1\over 2 m_N^2}{\partial v_S^+(r)\over r\partial r}\vec S
\cdot \vec L, \label{e:scex}
\end{equation}
where $v_S^+(r)$ is a scalar function. The $\vec p\,^2/m^2$
term in the spin-independent part of this interaction
may be combined with the kinetic energy term in the
nuclear Hamiltonian, by replacing the nucleon mass
by the effective ``mass operator''
\begin{equation}
m^*(r)=m_N\left[1+{v_S^+(r)\over m_N}\right].
\end{equation}
To first order in $v_S^+(r)$, the scalar component of the
nucleon-nucleon interaction therefore implies the following
two-body ``correction'' to the single nucleon axial charge 
operator in Eq.~(2):
\begin{equation}
A_0^a(S)=-{v_S^+(r)\over m_N} A_0^a(1) \pm (1\rightarrow 2)
.
\end{equation}
This interaction current operator coincides in form with
the ``scalar exchange'' pair current operator derived
in Refs. \cite{LeeR,Mariana}. 
The corresponding momentum space expression is
\begin{equation}
A_0^a(S)=-f_\pi {f_{\pi NN}\over m_\pi}
 {v_S^+ (k_2)\over m_N}
\vec \sigma^1\cdot\vec v_1 \tau_a^1
+(1\leftrightarrow 2), \label{e:pair}
\end{equation}
where $v_S^+(k)$ is the Fourier transform of the scalar
potential $v_S^+(r)$ and $\vec v_1 = (\vec p_1 +\vec p_1\,')/
2 m_N$ .
It is completely determined
by the nucleon-nucleon interaction 
model. Note that because the
scalar exchange interaction is attractive in realistic
nucleon-nucleon potentials, this exchange
current contribution implies an enhancement of the
cross section over the value given by the
single nucleon pion production mechanism. 

The expression for the vector exchange contribution to the
axial charge operator as derived from the 5-point
function has been given in Refs.~\cite{LeeR,Mariana}:
\begin{equation}
A_0^a(V)=f_\pi {f_{\pi NN}\over m_\pi}
 {v_V^+(k_2)\over m_N}\left(
\vec\sigma^1\cdot \vec v_2+{i\over 2 m_N}
\vec\sigma^1\times\vec\sigma^2\cdot \vec k_2\right) \, . \label{e:pairvec}
\end{equation}
Here $v_V^+(k)$ is the isospin independent 
vector component of the nucleon-nucleon
interaction.

The axial charge operator is unique among the nuclear
current operators in that its short-range vector and
scalar exchange contributions add coherently rather
than cancel.

A large - and the longest range - fraction of the effective 
scalar and vector
exchange components of the nucleon-nucleon
interaction is due to two-pion exchange. A direct
calculation of this uncorrelated part
of the two-pion exchange component
has recently been attempted by means of ChPT in ref.\cite{Kubo}. 
The results of that calculation indicate that the
two-pion exchange mechanisms are large, although the
largest contribution arises from a hybrid $\pi-2\pi$
exchange diagram.

%\vspace{1cm}
%\centerline {\bf 3. Nucleon resonance contributions}
%\vspace{0.5cm}
\section{Nucleon resonance contributions}
%==================================================================

%\centerline{\it 3.a The N(1440) contribution}
%\vspace{0.5cm}
\subsection{The N(1440) contribution}

The nucleon resonance contributions to pion production
in nucleon-nucleon collisions are illustrated by the
Feynman diagrams in Fig.~\ref{Fig.2}.
The N(1440) is the lowest vibrational state of the nucleon,
and as such should be excited by the same exchange 
mechanisms that appear in the nucleon-nucleon interaction.
In particular it is expected to be excited by the effective
scalar and vector fields in the nucleon  \cite{Pena}.

The contribution to the amplitude for the reaction
$pp\rightarrow pp\pi^0$ from virtual intermediate
$N(1440)$ resonances excited by effective isospin
independent scalar exchange mechanisms may be written
in terms of the corresponding amplitude for the scalar 
exchange contribution implied by the nucleon-nucleon
interaction (\ref{e:pair}) as
\begin{equation}
A_0^a(S,N(1440))=K_\sigma A_0^a(S),\label{e:A0}
\end{equation}
where the coefficient $K_\sigma$ is defined as
\begin{equation}
K_\sigma={g_{\sigma NN^*}^{1440}\over g_{\sigma NN}}
{f_{\pi NN^*}^{1440}\over
f_{\pi NN}}{2m_N\over m_{N^*}^{1440}-m_N},
\end{equation}
in the sharp resonance approximation.
Here the $f_{\pi NN^*}^{1440}$ denotes the pseudovector
coupling constant for the $\pi N N(1440)$ vertex, and
$g_{\sigma NN}$ and $g_{\sigma N N^*}^{1440}$ the
coupling strengths for an effective scalar meson,
the exchange of which represents the effective
scalar field (mainly correlated two-pion) exchange
interaction between the nucleons.

        The pseudovector $\pi NN(1440)$ coupling
constant is determined by the experimental
partial decay width for $N(1440)\rightarrow \pi N$ for
all accessible charge channels as
\begin{equation}
{(f_{\pi NN^*}^{1440})^2\over 4\pi}
= {1\over 3}{m_\pi^2 m_{N^*}^{1440}\over p(m_{N^*}^{1440}+m_N)^2
(E_N-m_N)}\Gamma
\left[N(1440)\rightarrow \pi N\right] \, , \label{e:width}
\end{equation}
which results in $(f_{\pi NN^*}^{1440})^2/4\pi\simeq 0.01$.
This value is close to that in \cite{Garz}, and smaller
by a factor 3 than that used in ref.
 \cite{Pena}, the difference
arising from the inclusion of all different charge
states in the decay width calculation. 
In (\ref{e:width}), $p$ and $E_N$ are the momentum
and energy of the final nucleon, respectively.

The numerical value for the coupling constant $g_{\sigma NN}$ is
determined by the nucleon-nucleon interaction model.
The value of the $N(1440)$ coupling to the scalar
field is very uncertain  \cite{Mad}. It was determined
from the empirical fractional decay for
$N(1440)\rightarrow N(\pi\pi)_{S wave}^{I=0}$ in
 \cite{Pena} as
\begin{equation}
{(g_{\sigma NN^*}^{1440})^2\over 4\pi}={m_{N^*}^{1440}        
\over p(E_N+m_N)}\Gamma\left[N(1440)\rightarrow N(\pi\pi)_
{S wave}^{I=0}\right] \, ,
\end{equation}
yielding a numerical value of 
$(g_{\sigma NN^*}^{1440})^2 /4\pi\simeq 0.1$
for $\Gamma[N(1440)\rightarrow N(\pi\pi)_
{S wave}^{I=0}]=35$ MeV. This value will be used here. 
In principle, measurement of photoproduction of vector
mesons and the $N(1440)$ resonance off protons should
provide fairly direct information on the magnitude
of this constant. With the value above we obtain for the
ratio $g_{\sigma NN^*}^{1440}/g_{\sigma NN}$ the value
0.11, as $g^2_{\sigma NN}/4\pi = 8.28$ in the
Bonn B potential model  \cite{Machl}.

The intermediate $N(1440)$ resonance may also be excited by
$\omega-$meson exchange. This contribution separates into
one that arises from the charge coupling of the $\omega-$meson
and one that arises from the current couplings  \cite{VanR}.
This term may be expressed as an axial charge operator of the
form (in momentum space)
\begin{eqnarray}
\lefteqn{
A_0^a(V)=f_\pi {f_{\pi NN}^*\over m_\pi}
\left({g_{\omega NN^*}^{1440}\over
g_{\omega NN}}\right){4 m_N\over (m_{N^*}^{1440})^2-m_N^2}
v_V^+(k_2)
}
\nonumber \\
& & \times 
\left[
\vec\sigma^1\cdot \vec v_1+{m_{N^*}^{1440}-m_N
\over 2 m_N}\vec\sigma^1\cdot\left(\vec v_2+
{i\over 2m_N}\vec\sigma^2\times\vec k_2\right)\right]\tau^1_a
+ (1\leftrightarrow 2) \, . \label{e:roper}
\end{eqnarray}
Here, $v_V^+(k)$ is the coefficient of the isospin 
independent vector exchange component of the nucleon-nucleon
interaction, and $g_{\omega NN^*}^{1440}$
and $g_{\omega NN}$ 
are the $\omega NN(1440)$ and $\omega-$nucleon 
coupling
constants, respectively. 
The momentum transfer to nucleon 2 is denoted $\vec k_2$.
Following Ref.~\cite{Pena},
we assume that the ratio between the scalar and vector
coupling constants are the same for the $N(1440)N$ couplings
as for the diagonal nucleon couplings. Thus we take
$g_{\omega NN^*}^{1440}/g_{\sigma NN^*}^{1440}
=g_{\omega NN}/g_{\sigma NN}=1.55$.

It is obvious from the expression (\ref{e:roper}) that the first term
on the rhs, which is that arising from charge coupling,
goes against that of the scalar exchange contribution (\ref{e:pair}).
This suggests that they should be considered together. 
The resulting partial cancellation makes the net
contribution of the intermediate $N(1440)$ resonance 
small.

%\vspace{1cm}
%\centerline{\it 3.b The N(1535) contribution}
%\vspace{0.5cm}
%==================================================================
\subsection{The N(1535) contribution}

The $N(1535)$ resonance stands out by its exceptionally
large $N\eta$ decay width, given its closeness to 
the threshold for $\eta-$decay. As a consequence, the
$\eta NN(1535)$ coupling constant has to be very large,
and therefore the $N(1535)$ excitation by $\eta-$meson
exchange contributes a nonnegligible amount to the
cross section for $pp\rightarrow pp\pi^0$. 

To describe this contribution, we describe the $\eta NN$ and
$\eta NN(1535)$ couplings by the Lagrangians
\begin{mathletters}
\begin{equation}
{\cal L}_{\eta NN}=i {f_{\eta NN}\over m_\eta}
\bar \psi\gamma_5\gamma_\mu
\partial_\mu \eta \psi, \label{e:etalaga}
\end{equation}
\begin{equation}
{\cal L}_{\eta NN(1535)}=i{f_{\eta NN^*}^{1535}\over m_\eta}
\bar\psi(1535)\gamma_\mu\partial_\mu \eta\psi+h.c. 
\label{e:etalagb}
\end{equation}
\end{mathletters}
The $\eta NN$ pseudovector coupling constant $f_{\eta NN}$
is contained in the potential model, and is related to the
corresponding pseudoscalar coupling constant as
$f_{\eta NN}=m_\eta g_{\eta NN}/2 m_N$. 

The coupling constant $f_{\eta NN^*}^{1535}$ may be
calculated from the partial width for $N\eta$ decay
of the $N(1535)$ as  \cite{Cheng}:
\begin{equation}
{(f_{\eta NN^*}^{1535})^2\over 4\pi}=
{m_\eta^2 m_{N^*}^{1535}\over p\omega_\eta^2( E_N+m_N)}
\Gamma\left[N(1535)
\rightarrow N\eta\right] \, .
\label{e:feta}
\end{equation}
The empirical partial decay width 67 MeV then yields the
value $(f_{\eta NN^*}^{1535})^2/ 4\pi=0.24$.

The $\pi N N(1535)$ coupling has the same form as (\ref{e:etalaga}),
with the modification that $\eta$ is replaced by
$\vec\tau\cdot \vec\phi$, $m_\eta$ by $m_\pi$
and $\omega_\eta$ by $\omega_\pi$. 
The $\pi NN(1535)$ coupling
constant may then be calculated from the partial
width for $N\pi$ decay of the $N(1535)$ in analogy with
(\ref{e:feta}). From the empirical decay width 67.5 MeV we then
obtain the value
$(f_{\pi NN^*}^{1535})^2/ 4\pi=0.0021$. The smallness of this value has
been explained by chiral symmetry arguments 
 \cite{Oka}.

The $N(1535)$ resonance excited by $\eta$ exchange gives
rise to a contribution to the amplitude for $S-$wave
pion production in the reaction $pp\rightarrow pp\pi^0$,
which - to lowest order in the sharp resonance approximation,
and with neglect of delta function terms - 
may be expressed as an axial exchange charge
operator:
\begin{equation}
A_0^a(\eta)=-f_\pi {\omega_\pi f_{\pi NN^*}^{1535}
\over  m_\pi m_\eta^2}
{f_{\eta NN^*}^{1535}\over f_{\eta NN}}
{\vec\sigma^2\cdot \vec k_2 \vec\tau^1_a\over
m_{N^*}^{1535}-m_N}\left({m_\eta\over 2 m_N}\right)^2 v_{PS}^+(k_2)
+(1\leftrightarrow 2).
\end{equation}
Here, the function $v_{PS}^+$ is the isospin independent
pseudoscalar exchange component of the nucleon-nucleon 
interaction, which in a one-boson exchange
interaction model has the form
\begin{equation}
v_{PS}^+(k)={g_{\eta NN}^2\over m_\eta ^2+ \vec k^2}
,
\end{equation}
with $g_\eta = 2m_N f_{\eta NN}/m_\eta$.

%\vspace{1cm}
%\centerline{\it 3.c The N(1520) contribution}
%\vspace{0.5cm}
%==================================================================
\subsection{The N(1520) contribution}
 
The $N(1520)$ is the spin $3/2^-$ partner of the $N(1535)$.
Although almost degenerate in mass, its structure - as
indicated by its decay pattern - is quite different.
The outstanding feature is the substantial decay
branch (10-15\%) to the $N\rho$ channel. As this
decay kinematically can reach only the lower end of 
the $\rho$ meson spectrum, the corresponding coupling
constant has to be very large. Therefore, the
$N(1520)$ resonance excited by $\rho-$meson exchange should
be expected to give a non-negligible contribution 
to the reaction $pp\rightarrow pp\pi^0$.

The coupling of pions and $\rho-$mesons to the 
$N(1520)$ resonance may be described by the 
Lagrangians
\begin{mathletters}
\begin{equation}
{\cal L}_{\pi NN(1520)}=i{f_{\pi NN^*}^{1520}\over m_\pi}  
\bar \psi_\mu(1520) \gamma_5\vec \tau\cdot\partial_\mu\vec\phi
\psi+h.c.,
\end{equation}
\begin{equation}
{\cal L}_{\rho NN(1520)}=ig_{\rho N N^*}^{1520}
\bar\psi_\mu(1520)
\left(\delta_{\lambda\mu}+{\partial_\mu\over m_{N^*}^{1520}
-m_N}\gamma_\lambda\right)\vec\rho_\lambda\cdot\vec\tau
\psi+h.c. \,\, . \label{e:lagrho}
\end{equation}
\end{mathletters}
The second term in (\ref{e:lagrho}) is required by the transversality
of the $\rho-$field.

The $\pi N N(1520)$ coupling constant may be determined
from the partial width for the decay $N(1520)
\rightarrow N\pi$ to all charge channels as  \cite{Cheng}:
\begin{equation}
\frac{(f_{\pi NN^*}^{1520})^2}{4\pi}={m_{N^*}^{1520} m_\pi^2
\over p^3 (E_N-m_N)}\Gamma\left[N(1520)
\rightarrow N\pi\right] \, .
\end{equation}
This yields $(f_{\pi NN^*}^{1520})^2/4\pi=0.19$ and
$f_{\pi NN^*}^{1520}=1.6$.

The $\rho NN(1520)$ coupling constant may be estimated from the
partial width for $\rho N$ decay of the $N(1520)$, which is
about 24 MeV  \cite{PDG}. To lowest order in $p/m_N$, this
decay rate may be written as
\begin{equation}
\Gamma\left[N(1520)\rightarrow N\rho\right]={2\over 9}
{(g_{\rho NN^*}^{1520})^2\over 4\pi}{p\over m_{N^*}^{1520}}
{m_N\over E_N}(E_N+m_N). \label{e:rhowidth}
\end{equation}
This expression, however, only applies for kinematically
allowed decays. In the case of $\rho$ decay of the
$N(1520)$ only the lower tail of the $\rho-$meson spectrum
between $2m_\pi$ and the kinematical threshold at 581 MeV
is accessible. If the $\rho-$meson spectrum is described
by a Lorentzian centered at 770 MeV with a full width of
151 MeV  \cite{PDG}, the weight of the kinematically
allowed part of the spectrum for $\rho-$meson decay is
only 0.076. That factor should then be included on the
rhs of Eq.~(\ref{e:rhowidth}). 

To determine the kinematical factors $p$ and $E_N$ in (\ref{e:rhowidth}),
a $\rho-$meson mass has to be specified. This will be taken
to be 480 MeV, which is the average of the kinematically
allowed mass value range, weighted with the $\rho-$meson
spectrum. Using this mass value, and taking into account
the probability factor 0.076, Eq.~(\ref{e:rhowidth}) gives the values 
$(g_{\rho NN^*}^{1520})^2/4\pi=0.43$, 
and $g_{\rho NN^*}^{1520}=2.3$.
This magnitude of this 
coupling constant is only about one half that (-5.3) 
used in Ref.~\cite{Peters} (The decay width does not
determine the sign of this coupling constant).

The isospin symmetric part
of the contribution of the intermediate $N(1520)$ resonance
excited by $\rho-$meson exchange to the amplitude for
$S-$wave pion production may be expressed -- to lowest order in 
$1/m$ -- as an axial
exchange charge operator with the form
\begin{eqnarray}
A_0^a(\rho) & = & f_\pi{2\over 3}{f_{\pi NN^*}^{1520}
\over m_\pi( m_{N^*}^{1520}-m_N)}
\left({g_{\rho NN^*}^{1520}\over g_{\rho NN}}\right)v_V^-(k_2)
\vec \sigma^1\cdot \left[\left(
{m_N\over  (m_{N^*}^{1520})^2-m_N^2}\right)
\vec k_2 \right.
\nonumber \\
& &
\left. 
-{2m_N\over m_{N^*}^{1520}+m_N}\vec v_1
+ \vec v_2+{i\kappa_\rho
\over 2m_N}
\left(\vec\sigma^2\times\vec k_2\right)\right]
\tau_a^2
 +(1\leftrightarrow 2).
 \end{eqnarray}
Here, the function $v_V^-(k)$ is the isospin dependent part
of the vector exchange component of the nucleon-nucleon
interaction, which in a boson exchange model would have the
form
\begin{equation}
v_V^-(k)={g_{\rho NN}^2\over m_\rho^2+\vec k^2} \, .
\end{equation}
The parameter $\kappa_\rho$ is the $\rho NN$ tensor
coupling constant.
With the positive sign for all resonance coupling
constants, this contribution would have the same sign as that
of the $N(1440)$ resonance. The data, however, favor the
negative value for the $\rho NN(1520)$ coupling (as used in
Ref.~\cite{Peters}), as a consequence of which 
its contribution
opposes the effect of the N(1440).

%\vspace{1cm}
%\centerline{\bf 4. Numerical Results }
%\vspace{0.5cm}
%==================================================================
\section{Numerical Results}

The calculated values of the total cross section for the
reaction $pp\rightarrow pp\pi^0$ are shown in Fig.~\ref{Fig.3} for
energies near threshold. The {\em cumulative}
contributions of the
different components in the transition amplitude are
shown successively.

The short-dashed curve is that obtained with the single nucleon
axial current operator (2) combined with the pion 
exchange term (3). This result confirms the calculations
of Refs. \cite{Myhr,VanK} that there is a strong
destructive interference between the impulse and
rescattering terms.
The pion exchange
contribution was calculated with full account of the
energy of the exchanged pion as in Ref. \cite{Myhr2}.

The short-dash-dotted curve in Fig.~\ref{Fig.3} includes the
short-range contributions associated with the scalar
(\ref{e:pair}) and vector exchange (\ref{e:pairvec})
contributions to the nucleon-nucleon interaction (``Z-graphs'').
Here the
potential coefficients $v_S^+$ and $v_V^+$ were
taken from the Bonn B potential \cite{Machl} by the
method described in Ref.~\cite{Pena}. When using this potential
model, the cumulative cross section obtained with
inclusion of these short-range effects are close
to the empirical values given in Refs. \cite{Meyer1,Meyer2,Celsius}.
This result also differs from those obtained
earlier with other potential models, where additional short-range
mechanisms had to be included in order to reach the empirical
cross section \cite{VanR}.

Turning then to the contributions of the
nucleon resonances, it is first noted that the result
for the $N(1440)$ resonance (dashed curve) 
is sizable, when calculated with the
coupling constants used in Ref.~\cite{Pena}. Since
the value for the $\sigma NN(1440)$ coupling constant
could still be significantly larger \cite{Mad,Hir,Ose}, this 
contribution could, accordingly, be even more pronounced, and the 
result reported here may be interpreted as a lower limit.
The resonance $N(1535)$ (short-dotted) has a negligible effect and 
the corresponding curve almost coincides with the dashed curve from
the $N(1440)$ resonance. The
$N(1520)$ (if the $\rho NN(1520)$ coupling constant
is negative), 
as can be seen from the dashed and the solid curves,  
counteracts the effect of the Roper resonance $N(1440)$,
and therefore the net effect of these resonances is small.

When all resonances are included, the calculated cross
section exceeds the empirical values at the
higher end of the energy range considered. This feature
suggests that there should be further short range contributions
(c.f. \cite{VanR}) that may play a role. In ref. \cite{LeeR},
for instance, the small contributions from the isospin I=1
scalar and vector exchanges were considered.  In addition, the
relativistic corrections to the rescattering amplitudes 
will become more significant with increasing pion momentum.

Since the signs of the mesonic resonance transition 
couplings are not fixed by the empirical
partial widths and decay rates, we show
in Fig.~\ref{Fig.4} the uncertainty implied by the 
uncertainty in sign and magnitude of the $\rho NN(1520)$
coupling in the calculated cross section. 
The figure indicates that with positive signs for the
couplings of the other resonances, the negative sign 
for the  $\rho NN(1520)$ vertex coupling is favored.

In order to illustrate the sensitivity of the calculated 
cross section to the details of the dynamics and the
conventional approximations, 
we also show in Figs.~\ref{Fig.5} and \ref{Fig.6} 
how the results for the impulse and the pion exchange
terms change if one uses the so-called ``frozen kinematics approximation''
applied to those operators. In this approximation, the production
operators are evaluated in exact threshold kinematics, even for 
energies above threshold.
The calculation of the cross-section
is simplified in this case, since the pion production amplitude does
not depend on the pion momentum variable of integration.  

Clearly, using frozen kinematics is not a good approximation to the exact
calculation in either case, neither for the impulse term of 
Fig.~\ref{Fig.5}, nor
for the pion re-scattering term of Fig.~\ref{Fig.6},
in agreement with ref. \cite{Myhr2}. The discrepancy -- not
surprisingly -- worsens the further one goes from threshold. 

We also consider the approximation that the energy dependence
of the 
half-off-shell T-matrix, in the final state, is replaced by the energy
dependence of 
the on-shell T-matrix \cite{Cebaf}. It is based on the observation
that, although the 
on-shell and half-off-shell T-matrix elements entering in the
calculation of pion production can be very different, their variation with 
energy is very similar and can be isolated into a factor dominated by the
two-nucleon phase shift. It suffices then to calculate the half-off-shell 
T-matrix at only one energy, e.g., at threshold, and the factor containing 
the phase shift takes care of the extension to all other energies.
This simplifies the phase-space integration, without sacrificing
the final state interaction. 

Figures ~\ref{Fig.5} and \ref{Fig.6} display the effect of this
approximation 
in the following way:  the frozen kinematics approximation has
been calculated on one 
hand using the exact half-off-shell two-nucleon T-matrix
(dotted curve),
and on the other 
hand with the described energy-dependence factorization (dashed
curve).
Therefore, the 
dotted curve has to be compared with the dashed curve in
order to asses the 
quality of this approximation, {\em not} with the solid line
representing the result without any of the considered approximations.

The energy-dependence approximation works somewhat better than the frozen
kinematics approximation. As to be expected, the discrepancy 
increases with the distance
in energy to the 
point at which the half-off-shell T-matrix has been calculated
exactly -- in this case at threshold. 

The frozen kinematics calculation overestimates the impulse term, but
underestimates the pion re-scattering term. It is likely that partial
cancellations will 
occur when the two terms are calculated together, resulting
in a smaller discrepancy. However,
accidental cancellations of errors should not be interpreted in favor of an
approximation. It is 
quite clear from these results that the frozen kinematics 
approximation should be avoided.

%\vspace{1cm}
%\centerline{\bf 5. Conclusions }
%\vspace{0.5cm}
%==================================================================
\section{Conclusions}

The results obtained here indicate that the net effect of 
the orbitally excited intermediate nucleon resonances with 
mass below 1.6 GeV, that are excited by short-range
exchange mechanisms, to the calculated cross section for the
reaction $pp \rightarrow pp \pi^0$
is small. The resonance coupling parameters were all
determined phenomenologically from 
experimental decay widths and branching ratios.
The individual resonance contributions are not small, however,
and the size of the net effect depends on the sign of the 
transition coupling constants. The data seem to favor a negative
$\rho NN$(1520) coupling constant.

The results also support the view that the dominant dynamical
effects in the reaction $pp\rightarrow pp\pi^0$ are
(a) the single nucleon operator (2) \cite{Miller}, (b) the
short-range exchange mechanisms incorporated in the
short-range part of realistic nucleon interactions
(\ref{e:pair}) and (\ref{e:pairvec})
 \cite{LeeR}, and
(c) the pion exchange term (3). Of these it is presently
the last one that is most contentious  \cite{Myhr,VanK,Myhr2}
and which calls for further study. Additional
short-range contributions associated with
transition couplings between heavy mesons are
possibly also important, and interesting because
of their other phenomenological implications  \cite{VanR}.

The short-range exchange mechanisms associated
with the short-range components of the nucleon-nucleon
interaction may be derived directly from the
nonrelativistic limit of the nonsingular part of the
axial current 5-point functions with external leg
couplings, and are completely determined
by the nucleon-nucleon interaction  \cite{LeeR,Mariana}. 
These short-range mechanisms were here determined
from the ``Bonn B'' boson exchange model for
the nucleon-nucleon
interaction model. 
Their magnitude is larger than what the ``power counting
rules''  of Chiral Perturbation Theory naively would
suggest, namely
that the resonance mechanisms  should be larger than
the re-scattering
term and the ``5-point contact'' terms  \cite{VanK}. 
This is a consequence of the large effective
momentum scale involved in the exchange mechanisms.

%\vspace{1cm}
%\centerline {\bf Acknowledgments}
%\vspace{0.5cm}
%==================================================================
\section*{Acknowledgments}

We thank Fred Myhrer and M. Soyeur for instructive
correspondence. This
work was supported in part by the Academy of Finland under 
contracts 34081, 43982, and by the Portuguese FCT under contracts
PRAXIS/P/FIS/10031/1998 and CERN/C/FIS/1213/98. 
%\newpage
%\vspace{1cm}
%\centerline{\bf References}
\vspace{0.5cm}

%==================================================================

%\centerline{\bf Figure captions}
%\noindent

\newpage

\begin{figure}
\caption{Pair term or "Z-graph" representation 
of the axial exchange charge operator
contributions to pion production.}
\label{Fig.1}
\end{figure}

\begin{figure}
\caption{Nucleon resonance contributions to pion production
in nucleon-nucleon collisions.}
\label{Fig.2} 
\end{figure}

\begin{figure}
\caption{Total cross section as 
function of the proton lab energy. 
The meaning of the various curves is discussed in
the text.}
\label{Fig.3}
\end{figure}

\begin{figure}
\caption{Total cross section as 
function of the proton lab energy.
The solid curve corresponds to $g_{\rho NN^*}^{1520}=-2.3$,
the short-dotted curve to $g_{\rho NN^*}^{1520}=2.3$,
the dashed curve to $g_{\rho NN^*}^{1520}=-5.3$ and
the short-dash-dotted curve to $g_{\rho NN^*}^{1520}=-5.3$.}
\label{Fig.4}
\end{figure}

\begin{figure}
\caption{Total cross 
section based on exact and approximate calculations of the
single-nucleon term. The 
solid line is the exact result, the dotted line
applies the ``frozen 
kinematics approximation'' (q=0), and the dashed line refers to a
calculation where, in 
addition, the energy-dependence of the nucleon-nucleon T-matrix
is approximated, as discussed in the text.}
\label{Fig.5}
\end{figure}

\begin{figure}
\caption{Total cross 
section based on exact and approximate calculations of the
pion re-scattering term. The meaning of the various curves is explained in
Fig.5}
%Fig.~\ref{Fig.4}.}
\label{Fig.6}
\end{figure}


\begin{references}
\bibitem{LeeR} T.-S. H. Lee and D. O. Riska, Phys. Rev. Lett.
{\bf 70}, 2237 (1993).
\bibitem{Myhr} B. Y. Park et al., Phys. Rev. {\bf C53}, 1519 (1996). 
\bibitem{VanK} T. D. Cohen et al., Phys. Rev. {\bf C53}, 2661 (1996).
\bibitem{Meyer1} H. O. Meyer et al., Phys. Rev. Lett. {\bf 65}, 2846
(1990).
\bibitem{Meyer2} H. O. Meyer et al., Nucl. Phys. {\bf A539}, 633 
(1995).
\bibitem{Celsius} A. Bondar et al. Phys. Lett. {\bf B356}, 8 (1995).
\bibitem{Miller} G. A. Miller and P. U. Sauer, Phys. Rev. 
{\bf C44}, R1725 (1991).
\bibitem{Oset} E. Hern\'andez and E. Oset, Phys. Lett. {\bf B350},
158(1995).
\bibitem{Hanhart} C. Hanhart et al., Phys. Lett. {\bf B424},
8 (1998).
\bibitem{Tamura} K. Tamura, Talk presented at the INT/Argonne 
Mini-Workshop on Pion Production Near Threshold, August 12-14 (1998)
\bibitem{VanR} U. Van Kolck, G. A. Miller and D. O. Riska,
Phys. Lett. {\bf B388}, 679 (1996).
\bibitem{Nisk} J. Niskanen, Phys. Lett. {\bf B289}, 227 (1992).
\bibitem{Mad} M. Soyeur, Acta. Phys. Polonica {\bf B29}, 2501 (1998).
\bibitem{Machl} R. Machleidt, Advances in Nuclear and Particle
Physics, {19} 189 (1989).
\bibitem{Myhr2} T. Sato et al, Phys. Rev. {\bf C56}, 1246 (1997).
\bibitem{Weinberg} S. Weinberg, Phys. Rev. Lett. {\bf 18}, 188 (1967).
\bibitem{Nyman} E. M. Nyman, Phys. Lett. {\bf B215}, 29 (1988).
\bibitem{Kubodera} K. Kubodera and M. Rho, Phys. Rev. Lett.
{\bf 67}, 3479 (1991). 
\bibitem{Mariana} M. Kirchbach, D. O. Riska and K. Tsushima,
Nucl. Phys. {\bf A542}, 616 (1992).
\bibitem{Towner} I. Towner, Nucl. Phys. {\bf A542}, 631 (1992).
\bibitem{Moalem} E. Gedalin, A. Moalem and L. Razdolskaya, 
``$\chi$PT Calculations with Two-pion Loops
for S-wave $\pi^0$ Production in pp collisions'',
eprint nucl-th/9803029.
\bibitem{Meissner} V. Bernard, N. Kaiser and U.-G. Meissner,
Int. J. Mod. Phys. {\bf E4},  193 (1995).
\bibitem{Coe} F. Coester and D. O. Riska, Ann. Phys. (N. Y.)
{\bf 234}, 141(1994).
\bibitem{Kubo} V. Dmitrasinovic et al., eprint nucl-th/9902048
\bibitem{Pena} S. Coon, M. T. Pe\~na and D. O. Riska,
Phys. Rev. {\bf C52}, 2925 (1995).
\bibitem{Garz} H. Garcilazo and E. Moya de Guerra,
Nucl. Phys. {\bf A562}, 521 (1993)
\bibitem{Cheng} W. K. Cheng and C. W. Kim, Phys. Rev.
{\bf 154}, 1525 (1967).
\bibitem{Oka} D. Jido, M. Oka and A. Hosaka,
Phys. Rev. Lett. {\bf 80}, 448 (1998).
\bibitem{PDG} Particle Data Group, Phys. Rev. {\bf D50}, 1173 (1994).
\bibitem{Peters} W. Peters et al., Nucl. Phys. {A632}, 609 (1998).
\bibitem{Hir} S. Hirenzaki et al., Phys. Rev. {\bf C33}, 277 (1996).
\bibitem{Ose} E. Hern\'andez and E. Oset, eprint 
nucl-th/9808017.
\bibitem{Cebaf} J. Adam Jr., Alfred Stadler, M. T. Pe\~na, Franz Gross,
Phys. Lett. {\bf B407}, 97 (1997).
\end{references}
\end {document}
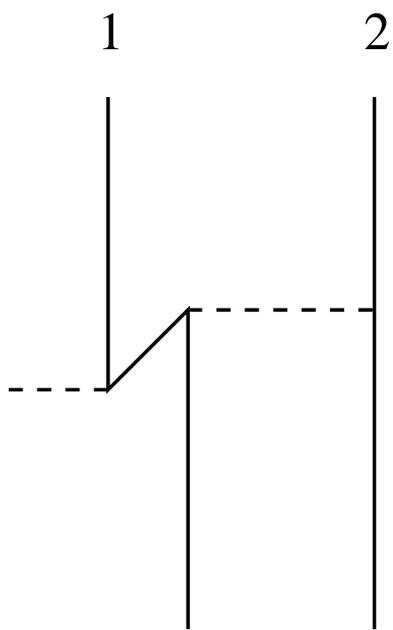
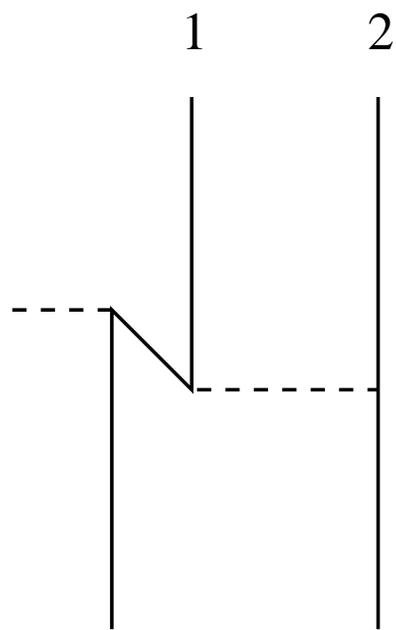
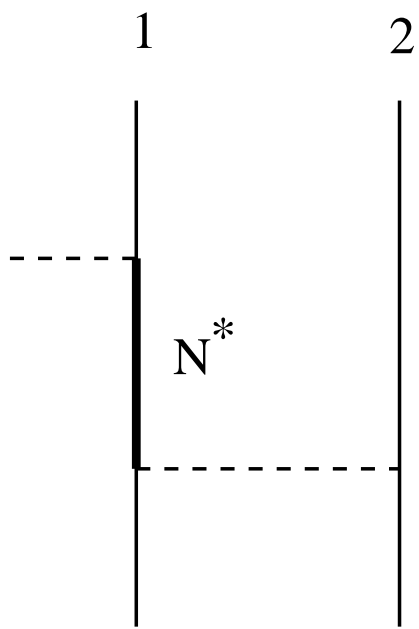
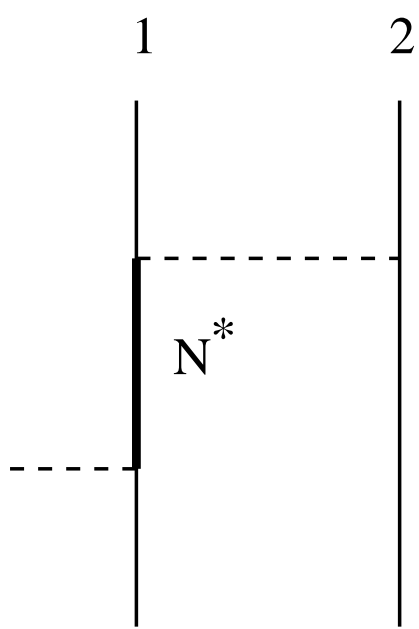
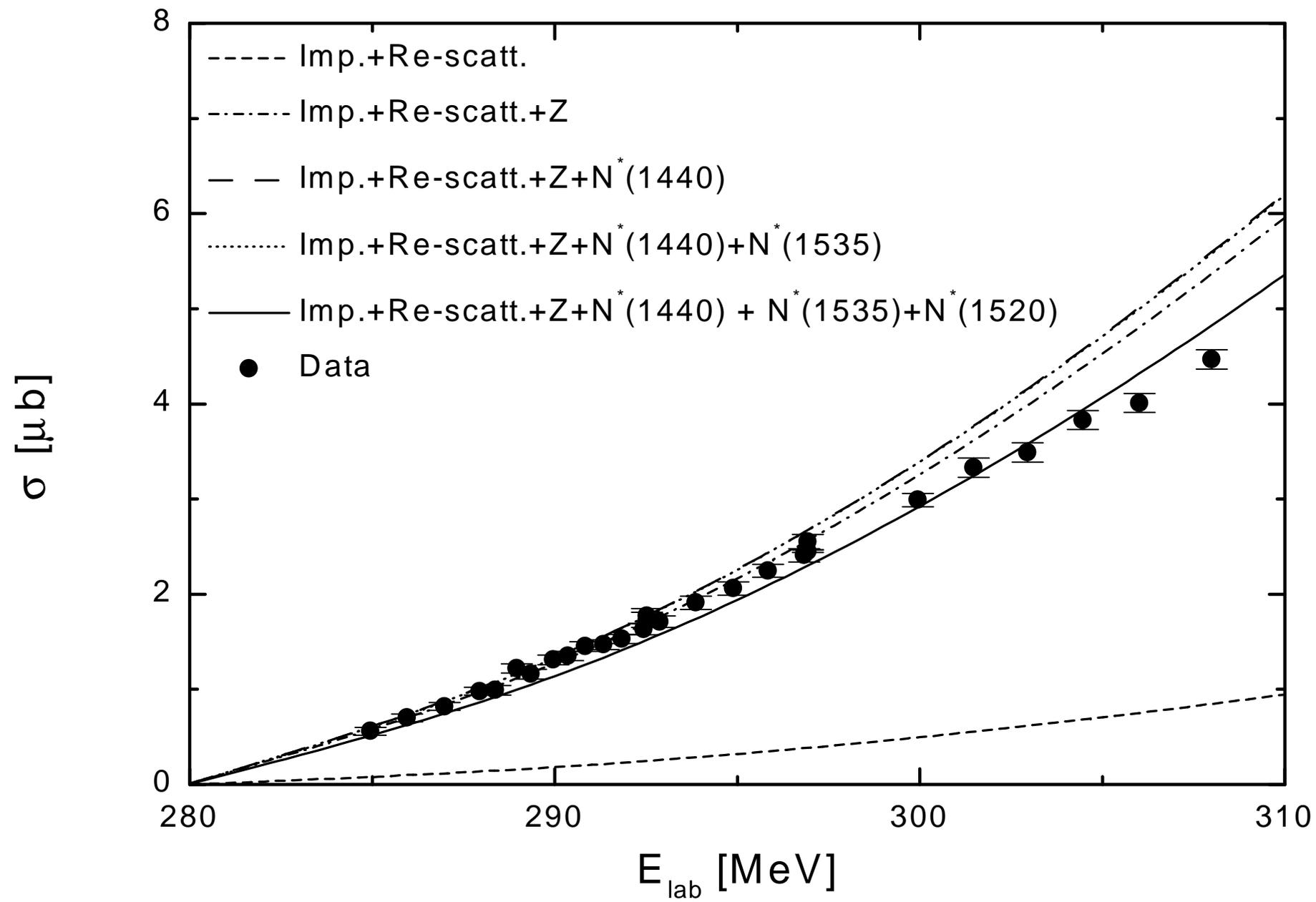
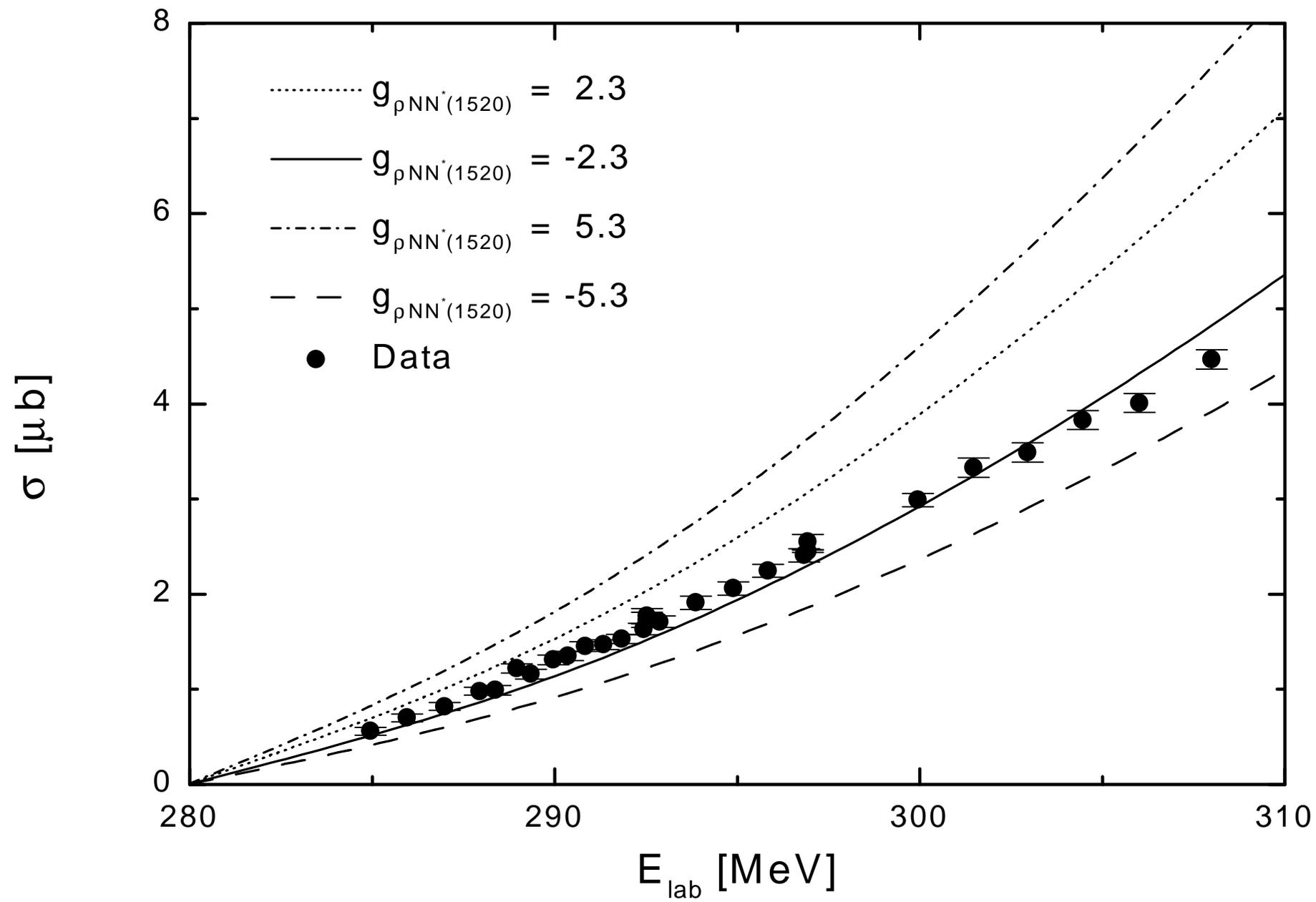
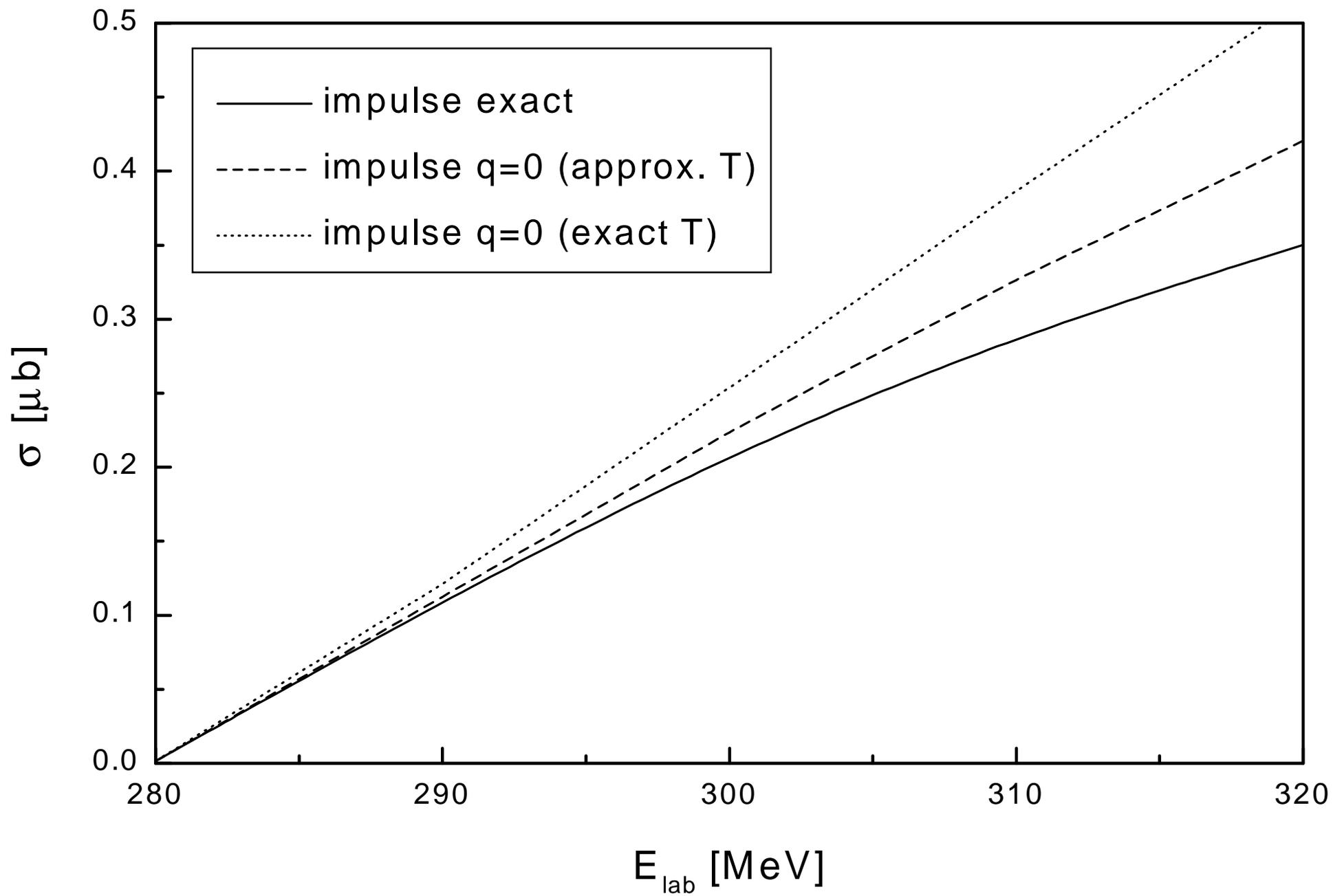
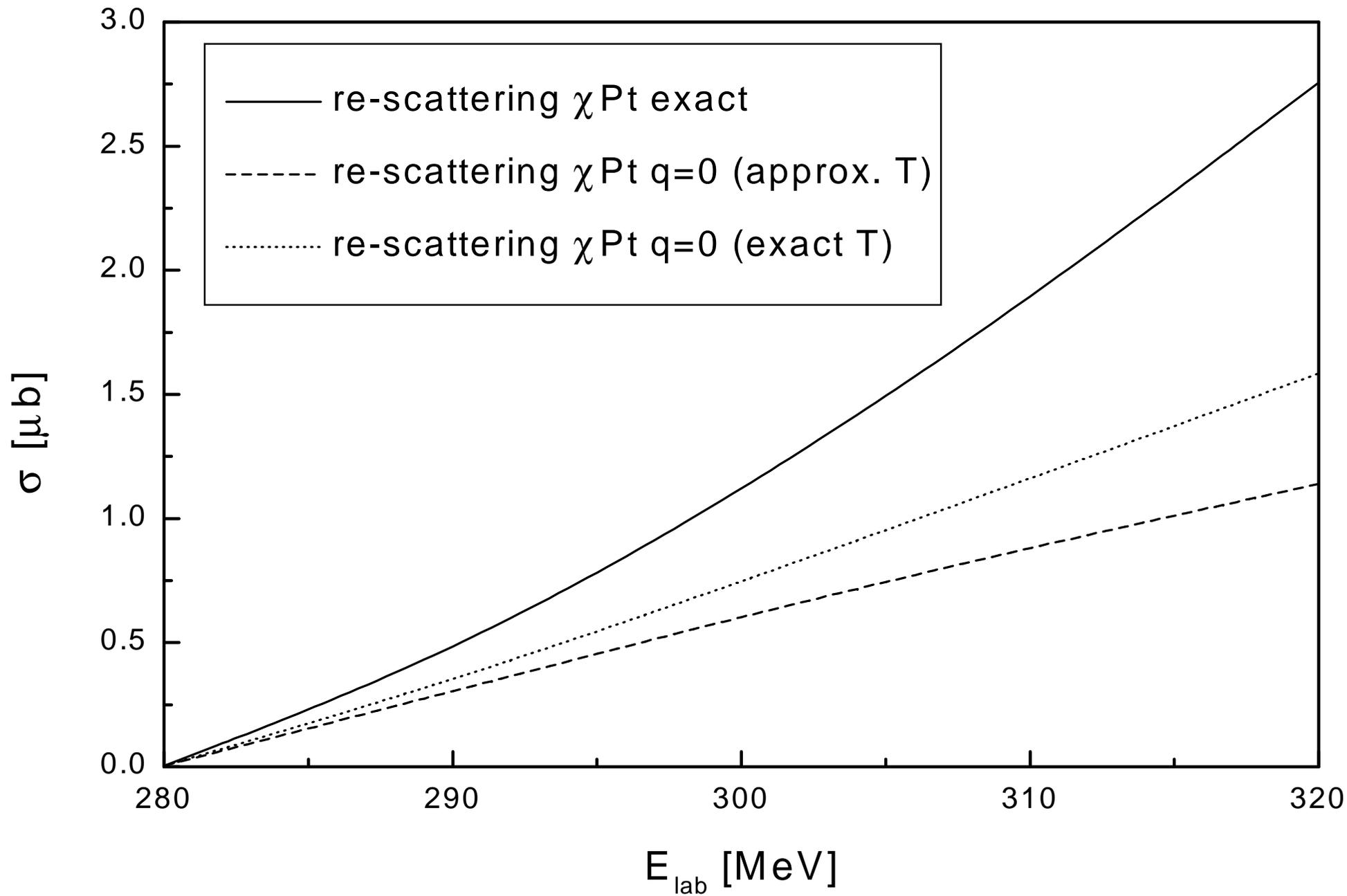
